\documentclass[twocolumn,pra,aps]{revtex4}
\usepackage{rotating}
\usepackage{hyperref}
\usepackage{graphicx}
\usepackage{amsmath}
\usepackage{amssymb}
\usepackage{mathtools}
\usepackage{color}
\definecolor{DarkBlue}{rgb}{0.1,0.1,0.55}
\definecolor{Blue}{rgb}{0,0,1}
\definecolor{Red}{rgb}{0.85,0.0,0.1}
\usepackage{epsfig}
\begin{document}
\bibliographystyle{/dane1/Chem-journal/jcp}
\title{\textit {Ab initio} studies of the ground and first excited states of the Sr--H$_2$ and Yb--H$_2$ complexes}
\author{Hubert Cybulski}
\email{hubert@fizyka.umk.pl, hcybulski@gmail.com}
\affiliation{Institute of Physics, Faculty of Physics, Astronomy and Informatics, Nicolaus Copernicus University in Torun, Grudziadzka 5, 87-100 Torun, Poland}

\begin{abstract}
Accurate intermolecular potential-energy surfaces (IPESs) for the ground and first excited states of the Sr-H$_2$ and Yb-H$_2$ complexes were calculated. After an extensive methodological study, the CCSD(T) method with the Douglas-Kroll-Hess Hamiltonian and correlation-consistent basis sets of triple-$\zeta$ quality extended with 2 sets of diffuse functions and a set of midbond functions were chosen.
The obtained ground-state IPESs are similar in both complexes, being relatively isotropic with two minima and two transition states (equivalent by symmetry). The global minima correspond to the collinear geometries with $R=$ 5.45 and 5.10~{\AA} and energies of $-$27.7 and $-$31.7~cm$^{-1}$ for the Sr-H$_2$ and Yb-H$_2$ systems, respectively.
The calculated surfaces for the Sr($^3P$)-H$_2$ and Yb($^3P$)-H$_2$ states are deeper and more anisotropic and they exhibit similar patterns within both complexes. The deepest surfaces, where the singly occupied \textit{p}-orbital of the metal atom is perpendicular to the intermolecular axis, are characterised by the global minima of ca. $-$2053 and $-$2260~cm$^{-1}$ in the T-shape geometries at $R=$ 2.41 and 2.29~{\AA} for Sr-H$_2$ and Yb-H$_2$, respectively.
Additional calculations for the complexes of Sr and Yb with the He atom revealed a similar, strong dependence of the interaction energy on the orientation of the \textit{p}-orbital in the the Sr($^3P$)-He and Yb($^3P$)-He states.
\end{abstract}
\maketitle
\newpage
\section{Introduction}
\label{intro}
Strontium (Sr) is an alkaline-earth metal with atomic number $Z=38$, while ytterbium (Yb) is a rare-earth metal with $Z=70$. However, because of its electronic configuration, Yb resembles the alkaline-earth elements. In both Sr and Yb the ground states of [Kr]~5\textit{s}$^2$ or [Xe]~4\textit{f}$^{14}$6\textit{s}$^2$ configurations, respectively, correspond to $^1S_0$ terms. This resemblance between Sr and Yb also implies similar schemes in their lowest excited states. The first excited triplet configurations for the Sr atom and the corresponding energies of the states are~\cite{NIST_ASD_21112018}: [Kr]~5\textit{s}5\textit{p} $\rightarrow$ $^3P_0$ (14~317.507~cm$^{-1}$), $^3P_1$ (14~504.334~cm$^{-1}$), and $^3P_2$ (14~898.545~cm$^{-1}$), while for the Yb atom we have: [Xe]~4\textit{f}$^{14}$6\textit{s}6\textit{p} $\rightarrow$ $^3P_0$ (17~288.439~cm$^{-1}$), $^3P_1$ (17~992.007~cm$^{-1}$), and $^3P_2$ (19~710.388~cm$^{-1}$). The spin-orbit coupling (SOC) splittings in Sr (186.831 and 394.212~cm$^{-1}$) and in Yb (703.568 and 1~718.381~cm$^{-1}$) demonstrate violation of the Land{\'e} interval rule~\cite{langhoff:84a}.

Simple forms of the ground-state interaction potentials with no hyperfine structure, availability of many stable isotopes,
the long-lived metastable $^3P^{\rm o}_0$ states resulting in the ultranarrow intercombination transitions of $^1S_0\rightarrow {^3P_0}$ at experimentally convenient wavelengths are the most important advantages of these two-electron elements, which made them attractive objects for research studies.
They have been successfully used in the fields of trapping ultracold quantum gas and Bose-Einstein condensation~\cite{takasu:2003a,takasu:2003b,martinez:2009a,stellmer:2009a,dareau:2015a}, quantum information processing~\cite{daley:2008a,gorshkov:2009a}, Fermi degeneracy~\cite{fukuhara:2007a}, studies of fundamental symmetries~\cite{demille:95a,tsigutkin:2009a} or photoassociation spectroscopy~\cite{takasu:2004a,tojo:2006a,enomoto:2007a}.

However, it seems that it is the optical atomic clock (see e.g. Refs.~\onlinecite{margolis:2009a} and~\onlinecite{ludlow:2015a}) that receives the most attention, and thus quickly became a hot research topic. Driven by progress in the fields of atomic, optical and quantum science, vast improvements in atomic clocks was made and they soon surpassed the accuracy of caesium microwave clocks~\cite{ludlow:2008a,bloom:2014a,huntemann:2016a,campbell:2017a}. Optical atomic clocks are already being used to test fundamental theories and in development of new definitions of time and frequency standards~\cite{baber:2006a,takamoto:2015a} heralding a revision of the International System of Units (SI)~\cite{targat:2013a}.
They are also crucial for a rapid development in technologies that support broadband communication networks, navigation with global positioning systems (GPS) or clock-based geodesy~\cite{grotti:2018a}.

In each of the aforementioned standards, the concept of the reference frequency we tightly relate to the resonance frequency of the unperturbed atom or ion. However, in practise such a situation is impossible to attain and often, instead of being eliminated, the influence of external factors are usually best minimised and meticulously included in the systematic error budget. Thus, the actual accuracy depends on the control we have over the perturbations that the system experiences.

One of these factors are collisions of residual-gas atoms or molecules with the clock atom resulting in a frequency shift of the transition-line center. Recent improvements in the control of the systematic corrections in optical atomic clocks towards the 10$^{-18}$ level and beyond, make the collisional frequency shift~\cite{gibble:2013a,vutha:2017a} an important contribution to the error budget~\cite{rosenband:2008a}.
The measurements of the partial pressures of the background gases in the vacuum chamber showed~\cite{dube:2013a} that the most abundant gas is molecular hydrogen (approx. 60\% of the pressure).

Therefore, in this study we focus on the interactions of the Sr and Yb atoms in their ground and first excited states with the ground-state H$_2$ molecule.
The intermolecular potential-energy surfaces (IPESs) are \textit{ab intio} calculated using highly accurate methods and large basis sets.
The knowledge of both the ground and first-excited states is crucial for estimations of the collisional frequency shifts, which are essential for ultimate clock performance. No such studies have been done yet and the interactions remained completely unknown.

The manuscript is organised as follows: In Sec.~\ref{sec:comp} the details of the quantum-chemical calculations are given. Then, in Sec.~\ref{sec:res} the results of the studies are presented: In Sec.~\ref{sec:res:metbas} we describe methodological aspects and the results of the basis set studies, while in Sec.~\ref{sec:res:pess} the details of the IPESs are reported. Finally, in Sec.~\ref{sec:sum} we summarize and conclude.

\section{Computational details}
\label{sec:comp}
The geometries of the systems were characterised by three variables: the H--H bond length $r_{\rm HH}$, the distance $R$ from the Sr (or Yb) atom to the center of the H$_2$ molecule, and the angle $\theta$ between the H$_2$ molecular axis and the line connecting the Sr (or Yb) atom with the center of H$_2$. $\theta=0$ corresponds to the Sr-H$_2$ (or Yb-H$_2$) collinear orientation, while $\theta=90^{\circ}$ denotes the T-shape geometry of the complex. Because of the symmetry of the system, only the values $0 \leq \theta \leq 90^{\circ}$ needed to be considered.
In all calculations the value of $\langle$\textit{r}$_{\rm HH}\rangle =$ 0.7666393~{\AA} representing the \mbox{H--H} distance averaged over the ground vibrational state of H$_2$~\cite{roy:87a,jankowski:2005a} was used.

The quantum chemical calculations for systems containing heavier atoms are usually very challenging. Not only because of a large number of electrons to correlate making the studies time- and resource-consuming, but also because of partially multiconfigurational character resulting from mixing of several close-lying states, non-negligible relativistic contributions and a lack of families of basis sets allowing for estimation of the complete basis set limit.

In this study we have employed the spin-restricted coupled cluster with single, double and non-iterative triple excitation [R-CCSD(T)] method as implemented in MOLPRO (2012.1~\cite{MOLPRO_brief:2012.1} version). At the moment, this approximation represents the most sophisticated treatment of electron correlation computationally feasible for these systems. The spin-restricted approach was used to avoid spin contamination. To account for relativistic effects, the calculations were performed using the second- and third-order Douglas-Kroll-Hess (DKH) Hamiltonian~\cite{wolf:2002a,reiher:2004a,reiher:2004b} in the case of the Sr-H$_2$ and Yb-H$_2$ complex, respectively. Since the Sr($^3P$) and Yb($^3P$) states are high-spin cases and they are well separated (mainly by symmetry constrains), the use of the single-determinant spin-restricted open-shell Hartree-Fock orbitals is justified~\cite{klos:2002a,atahan:2006a,hapka:2013a}. Usually, a very tight convergence threshold (1$\cdot$10$^{-12}$) in energy calculations was set. In several cases convergence to the desired excited state was enforced by a suitable rotation of the occupied Hartree-Fock orbitals. The frozen-core approximation was employed with the 4\textit{s}$^2$4\textit{p}$^6$5\textit{s}$^2$ (Sr) and 5\textit{s}$^2$5\textit{p}$^6$6\textit{s}$^2$4\textit{f}$^{14}$ (Yb) electrons correlated (see also Sec.~\ref{sec:res:metbas}).

In some test calculations we employed the ECP28MWB pseudopotential~\cite{dolg:1989a} (PP) for the Yb atom. In this approximation all the 44 remaining electrons of the system were correlated.

The interaction energies were corrected for the basis set superposition error using the counterpoise (CP) method~\cite{boys:70a}.

To find the optimal basis set for the IPES calculations, we have started our study with an analysis of the performance of several available basis sets. The tests were carried out for the computationally more demanding Yb atom. We mainly focused on correlation-consistent family of basis sets developed for scalar relativistic calculations. The cc-pVXZ-DK3 (X$=$D, T, Q) basis set~\cite{lu:2016a} for the Yb atom and the cc-pVXZ-DK (X$=$D, T, Q) basis set (original exponents taken from Ref.~\cite{dunning:89a} and recontracted in Ref.~\cite{jong:2001a}) for the H atoms were used. A set of these two bases equal in X in this study we denote as XZ-DK. To test the influence of the correlation of core electrons on the calculated interaction energies, we have also used the cc-pwCVTZ-DK3 basis set~\cite{lu:2016a} for the Yb atom as well as the uncontracted TZ-DK basis set (see Sec.~\ref{sec:res:metbas}).

Further, the large atomic natural orbital­-relativistic core-­correlated (ANO-­RCC) basis set~\cite{widmark:90a,roos:2008a} was also employed.

In the calculations with the ECP28MWB pseudopotential, the ECP28MWB-ANO~\cite{cao:2001a} basis set was employed for the Yb atom and Dunning's augmented standard aug-cc-pVXZ (X$=$D, T) basis~\cite{dunning:89a,woon:93a} (denoted as aXZ) for the H atoms. This set was also extended by adding respectively 4 and 2 \textit{h}- and \textit{i}-type functions using the 4 and 2 lowest exponents of the \textit{g}-type functions.

In the calculations for the Sr-H$_2$ system, the aug-cc-pVTZ-DK2 basis set~\cite{hill:2017a} for the Sr atom and the aug-cc-pVTZ-DK basis set~\cite{dunning:89a,jong:2001a} for the H atom, further denoted as aTZ-DK, were used.
 
In some cases we increased the number of diffuse functions using the even-tempered scheme implemented in MOLPRO~\cite{MOLPRO_brief:2012.1} (since denoted as ``even''). All the basis sets were further extended by a set of the 3\textit{s}3\textit{p}2\textit{d}2\textit{f}1\textit{g}1\textit{h} midbond functions denoted as 332211, with exponents~\cite{tao:92a} of 0.90, 0.30,
and 0.10 for the \textit{s} and \textit{p}, 0.60 and 0.20 for \textit{d} and \textit{f},
and 0.30 for \textit{g} and \textit{h} functions, that were placed in the middle of the van der Waals bond.

In the calculations for the complexes with helium, the cc-pVTZ-DK and aug-cc-pVTZ-DK bases~\cite{woon:94a,jong:2001a} for the He atom were used (see Sec.~\ref{sec:res:pess:he}).

To evaluate the interaction energy in the complete-basis-set (CBS) limit, we have estimated the correlation part employing the formula proposed by Halkier \textit{et al.}~\cite{halkier:98a} for the correlation consistent basis sets series:

\begin{equation}
\label{eq:cbs:corr}
E_{XY}^{\rm corr, \infty}=\frac{E_X^{\rm corr}X^3-E_Y^{\rm corr}Y^3}{X^3-Y^3},
\end{equation}
where $X, Y$ are cardinal numbers of the basis sets, and $E_X^{\rm corr}, E_Y^{\rm corr}$ are the calculated correlation parts of the interaction energies. Using this estimate, the total interaction energy in the CBS limit can be calculated as:

\begin{equation}
\label{eq:cbs:tot}
E_{XY}^{\rm tot, \infty}=E_{Y}^{\rm HF}+E_{XY}^{\rm corr, \infty},
\end{equation}
where $E_{Y}^{\rm HF}$ is the Hartree-Fock (uncorrelated) part, usually calculated using the larger basis set.

\section{Results}
\label{sec:res}
\subsection{Methodological aspects and basis set study}
\label{sec:res:metbas}
As the first step in our study, we decided to test the PP approximation. The CCSD(T) interaction energies calculated with the ECP28MWB PP for the Yb-H$_2$ collinear geometry are collected in Table~\ref{tab:en:pp}. As a reference, in the last column we have added the CBS limit value estimated at the CCSD(T)-DKH3 level with the TZ-DK and QZ-DK basis sets extended with the 332211 midbond set (\textit{vide infra}).

\begin{sidewaystable}
\centering \caption{The CCSD(T) interaction energy values (in cm$^{-1}$) for the ground state of the Yb-H$_2$ complex in the collinear geometry. The intermolecular distance values $R$ are given in angstroms. In all the calculations the 332211 midbond function set was used. See Sec.~\ref{sec:comp} for more details regarding the used basis functions (bfs)}
\label{tab:en:pp} \vspace{0.5cm}
\begin{tabular}{lrrrrrrr}
\hline\hline
Yb&&&&&&&\\
PP&ECP28MWB&ECP28MWB&ECP28MWB&ECP28MWB&ECP28MWB&ECP28MWB&\\
Basis set&ECP28MWB-ANO&ECP28MWB-ANO&ECP28MWB-ANO&ECP28MWB-ANO&ECP28MWB-ANO&ECP28MWB-ANO&CBS$_{\rm TZ/QZ}$\\
Additional bfs&&1 even&\textit{h} + 1 even&\textit{h}, \textit{i} + 1 even&2 even&1 even&\\
\hline
H&&&&&&&\\
Basis set&aDZ&aDZ&aDZ&aDZ&aDZ&aTZ&\\
\hline
Number of bfs&178&203&258&297&228&231&229/330\\
\hline
R&\multicolumn{7}{c}{Interaction energy}\\
\hline
4.80&-21.17&-21.56&-22.06&-22.15&-21.16&-22.56&-26.45\\
5.00&-27.01&-27.11&-27.49&-27.64&-26.75&-28.65&-30.70\\
5.20&-28.00&-27.99&-28.22&-28.40&-27.80&-29.81&-30.80\\
5.40&-26.72&-26.62&-26.72&-26.84&-26.67&-28.21&-28.71\\
5.60&-24.61&-24.39&-24.43&-24.49&-24.59&-25.39&-25.55\\
\hline
\end{tabular}
\end{sidewaystable}

The results in Table~\ref{tab:en:pp} indicate that the minimum of the interaction energy corresponds approximately to $R=5.20$~{\AA} and its position seems to be independent of the used basis set. An addition of more diffused functions (1 or 2 even-tempered) decreases slightly the absolute value of the interaction energy (from $-$28.00 to $-$27.80~cm$^{-1}$), while an extension of the basis set with functions with greater angular momentum (\textit{h}, \textit{i}) increases (in absolute value) the calculated interaction energy ($-$27.99 \textit{vs.} $-$28.40~cm$^{-1}$). A more pronounced difference corresponds to the change of the hydrogen basis set from aDZ-DK to aTZ-DK (from $-$27.99 to $-$29.81~cm$^{-1}$). However, it is clear that the interaction energy is still underestimated. In the proximity of the minimum the difference (in absolute value) is ca. 1.0~cm$^{-1}$ in comparison with the CBS estimate and is clearly distance-dependent decreasing while the $R$ separation increases.

As the next step we performed frozen-core CCSD(T) calculations using the DKH3 Hamiltonian. In the calculations we employed the XZ-DK basis sets extended by the 332211 midbond function set and the resulting interaction energies are shown in Table~\ref{tab:en:bas}. For most of the bases the minimum is located in the proximity of $R=5.20$~{\AA}. However, for some of them, the position of the minimum shifts to shorter intermolecular distances. As one can expect, the interaction energy increases (in absolute value) with the increasing size of the basis set ($-$23.92 and $-$29.72~cm$^{-1}$ for the DZ-DK and QZ-DK bases, respectively), and a greater change is observed for the DZ-DK to TZ-DK transition ($-$23.92 \textit{vs.} $-$27.66~cm$^{-1}$) than for the TZ-DK to QZ-DK one ($-$27.66~cm$^{-1}$ and $-$29.72~cm$^{-1}$, respectively).

\begin{sidewaystable}
\centering \caption{The CCSD(T) interaction energy values (in cm$^{-1}$) for the ground state of the Yb-H$_2$ complex in the collinear geometry. The intermolecular distance values $R$ are given in angstroms. In all the calculations the 332211 midbond function set was used. See Sec.~\ref{sec:comp} for more details regarding the used basis functions (bfs). CBS$_{\rm X/Y}$ stands for the CBS limit calculated with the XZ-DK and YZ-DK basis sets}
\label{tab:en:bas} \vspace{0.5cm}
\setlength{\tabcolsep}{1.5mm}
\begin{tabular}{l|rrr|rrrr|rrr|rrrr|rr}
\hline\hline
Basis set&\multicolumn{3}{c|}{DZ-DK}&\multicolumn{4}{c|}{TZ-DK}&\multicolumn{3}{c|}{QZ-DK}&\multicolumn{4}{c|}{uANO-RCC}&CBS$_{\rm D/T}$&CBS$_{\rm T/Q}$\\
Additional bfs&&1 even&2 even&&1 even&2 even&3 even&&1 even&2 even&&1 even&2 even&3 even&&\\
\hline
Number of bfs&150&183&216&229&283&337&391&330&411&492&441&509&577&645&150/229&229/330\\
\hline
R&\multicolumn{16}{c}{Interaction energy}\\
\hline
4.80&-17.57&-28.29&-28.78&-21.95&-26.62&-26.92&-27.02&-25.03&-27.35&-27.51&-27.12&-27.47&-27.58&-27.61&-24.63&-26.45\\
5.00&-22.87&-32.42&-33.07&-26.94&-31.06&-31.27&-31.33&-29.46&-31.52&-31.63&-31.38&-31.58&-31.65&-31.68&-29.38&-30.70\\
5.20&-23.92&-32.02&-33.04&-27.66&-31.26&-31.41&-31.43&-29.72&-31.54&-31.62&-31.45&-31.56&-31.61&-31.64&-29.85&-30.80\\
5.40&-22.73&-29.51&-30.73&-26.10&-29.20&-29.32&-29.32&-27.78&-29.37&-29.43&-29.32&-29.38&-29.41&-29.43&-28.02&-28.71\\
5.60&-20.51&-26.19&-27.38&-23.48&-26.13&-26.21&-26.22&-24.80&-26.22&-26.27&-26.20&-26.22&-26.25&-26.27&-25.11&-25.55\\
\hline
\end{tabular}
\end{sidewaystable}

Using these results we have estimated the total interaction energies in the CBS limit as defined in Eq.~(\ref{eq:cbs:tot}) with $E_{Y}^{\rm HF}$ calculated at the QZ-DK level. In the calculations the midbond functions were used and it was shown that they can be very effective when extrapolation techniques are used for the CCSD and CCSD(T) energies~\cite{jeziorska:2003a,patkowski:2012a}. In the vicinity of the minimum the estimated interaction energies are $-$29.85 and $-$30.80~cm$^{-1}$ in the DZ-DK/TZ-DK and TZ-DK/QZ-DK basis set pairs, respectively and, again, this difference decreases with the $R$ distance.

We have also tested the effect of an addition of diffuse functions on the interaction energy of the complex. The effect is largest for the smallest, DZ-DK basis set. After the addition of two sets of diffuse functions (``DZ-DK+2 even'' in Table~\ref{tab:en:bas}), the obtained value of $-$33.04~cm$^{-1}$ seems to be overestimated in comparison with the reference TZ-DK/QZ-DK CBS limit of $-$30.80~cm$^{-1}$. A similar extension of a basis set for TZ-DK and QZ-DK causes a much smaller change ($-$3.75 and $-$1.90~cm$^{-1}$, respectively). It indicates that these bases are more balanced than the DZ-DK one. An addition of the third set of diffuse functions in the case of the TZ-DK basis set lowers the interaction energy only by 0.02~cm$^{-1}$. It means that this basis set is already saturated with diffuse functions.

In the basis set study we also include another large basis set, namely ANO-RCC. The original coefficients were uncontracted (uANO-RCC) and the basis was further supplied with the 332211 set of midbond functions. The calculated interaction energies are lower than those obtained with the QZ-DK basis set and at $R=5.20$~{\AA} they are $-$31.45 and $-$29.72~cm$^{-1}$, respectively. The uANO-RCC basis yields interaction energies well comparable with the results of the QZ-DK basis extended with one set of diffuse functions (``QZ-DK+1 even'' in Table~\ref{tab:en:bas}). Consecutive additions of diffuse functions to the uANO-RCC basis set result in fast convergence of the energies. Within the range of the $R$ distances presented in Table~\ref{tab:en:bas}, the difference between the interaction energies obtained with the uANO-RCC basis set supplied with three and two sets of diffuse functions does not exceed 0.03~cm$^{-1}$. It is worth noticing that the interaction energies calculated with the uANO-RCC and QZ-DK basis sets converge towards the same limit upon sequential addition of sets of diffuse functions. For $R$ greater than $5.20$~{\AA}, the 
uANO-RCC and QZ-DK results obtained with, respectively, two or three sets of diffuse functions are essentially the same (\textit{cf.} the ``uANO-RCC+3 even'' and ``QZ-DK+2 even'' interaction energies in Table~\ref{tab:en:bas}). This means that these results can be regarded as other references aside from the CBS limits.

Complete tables containing the basis set study results (also for T-shape orientations of the Yb-H$_2$ complex) are included in Supplementary Material.

All the discussed above CC calculations have been done in the frozen-core regime. This choice was confirmed \textit{a priori} by the pattern in the Hartree-Fock orbital energies: the orbital energies of the electrons kept uncorrelated in consequent CC calculations lied significantly lower than those of the correlated electrons. Now we shall discuss the validity of this approximation in more detail. To do this we performed some test calculations employing the cc-pwCVTZ-DK3 and cc-pVTZ-DK basis sets for the Yb and H atoms, respectively, (denoted here as wCTZ-DK) extended with the 332211 midbond functions set. The results obtained within the frozen-core (the valence 6\textit{s}$^2$4\textit{f}$^{14}$ electrons correlated along with the so-called ``outer-core'' 5\textit{s}$^2$5\textit{p}$^6$ electrons), inner-core (also 4\textit{s}$^2$4\textit{p}$^6$4\textit{d}$^{10}$ electrons correlated), and all-electron approximations are shown in Table~\ref{tab:en:bas:core}. For comparison, we added the frozen-core interaction energies calculated with the TZ-DK and uncontracted TZ-DK (uTZ-DK) basis sets (in both cases enhanced with the 332211 midbond set), too. It is clear that the differences between the TZ-DK and uTZ-DK results can be noticeable. However, they depend on the $R$ distance, can change sign, and in the proximity of the minimum are only 0.04~cm$^{-1}$. A similar pattern is observed for the differences between the frozen-core TZ-DK and wCTZ-DK results. At $R=5.20$~{\AA} this difference is ca. 0.08~cm$^{-1}$. Surprisingly, correlation of the inner-core electrons results in an increase of the interaction energy by ca. 0.2~cm$^{-1}$ (in comparison with the frozen-core approximation) and is weakly dependent on the $R$ distance (the ``ic'' column in Table~\ref{tab:en:bas:core}). Finally, the interaction energies calculated in the all-electron approach are essentially the same as those calculated in the inner-core regime. In the last column of Table~\ref{tab:en:bas:core} we have also included the all-electron results obtained with the wCTZ basis set supplied with 2 additional sets of diffuse functions. This extension of the basis set has a similar effect (ca. $-$3.7~cm$^{-1}$ in the vicinity of the minimum) as the transition between the TZ-DK and TZ-DK+2 even basis sets (\textit{cf.} Table~\ref{tab:en:bas}).

\begin{sidewaystable}
\centering \caption{The CCSD(T) interaction energy values (in cm$^{-1}$) for the ground state of the Yb-H$_2$ complex in the collinear geometry. The intermolecular distance values $R$ are given in angstroms. In all the calculations the 332211 midbond function set was used. See Sec.~\ref{sec:comp} for more details regarding the used basis functions (bfs). ``fc'', ``ic'', and ``ae'' stand for the frozen-core, inner-core, and all-electron approximations, respectively}
\label{tab:en:bas:core} \vspace{0.5cm}
\setlength{\tabcolsep}{2mm}
\begin{tabular}{l|rr|rrrr}
\hline\hline
Basis set&TZ-DK&uTZ-DK&\multicolumn{4}{|c}{wCTZ-DK}\\
Additional bfs&&&&&&2 even\\
Approximation&fc&fc&fc&ic&ae&ae\\
\hline
Number of bfs&229&385&270&270&270&378\\
\hline
R&\multicolumn{6}{c}{Interaction energy}\\
\hline
4.80&-21.95&-22.29&-22.22&-22.01&-22.02&-26.82\\
5.00&-26.94&-27.11&-27.10&-26.89&-26.88&-31.09\\
5.20&-27.66&-27.70&-27.74&-27.53&-27.53&-31.19\\
5.40&-26.10&-26.03&-26.12&-25.93&-25.93&-29.10\\
5.60&-23.48&-23.37&-23.46&-23.29&-23.29&-26.01\\
\hline
\end{tabular}
\end{sidewaystable}

To sum up, the total effect of the inclusion of the correlation of the inner-core electrons is non-negligible but relatively small. Since the (frozen-core) CCSD(T) results obtained with the DKH Hamiltonian seem to be more reliable than the PP ones, we have decided to perform the calculations employing the former approximation. Because of its medium size and the results close to those obtained with the well saturated uANO-RCC+3 even+332211 basis set, the TZ-DK basis set extended with 2 sets of diffuse functions and the 332211 midbond set (denoted as TZ-DK+2 even+332211) will be used in the following studies. Some additional tests (not presented here) showed that the difference between the results obtained in this basis and the uANO-RCC+332211 ones in the lowest excited Yb($^3P$)-H$_2$ state are also small. For consistency's sake, the aug-cc-pVTZ-DK2 basis set~\cite{hill:2017a} for the Sr atom and the aug-cc-pVTZ-DK basis set~\cite{dunning:89a,jong:2001a} for the H atom both extended with 2 sets of diffuse functions were used in the case of Sr-H$_2$.

\subsection{IPESs}
\label{sec:res:pess}
\subsubsection{Sr-H$_2$ and Yb-H$_2$}
\label{sec:res:pess:h2}
\begin{figure}[ht]
\caption{Schematics of the singly occupied $p$-orbital orientation in the first excited state of the Me-H$_2$ (Me $=$ Sr,Yb) complexes for the collinear (a-c) and T-shape (d-f) configurations}
\label{fig:orbit}
{\scalebox{0.4}{\includegraphics{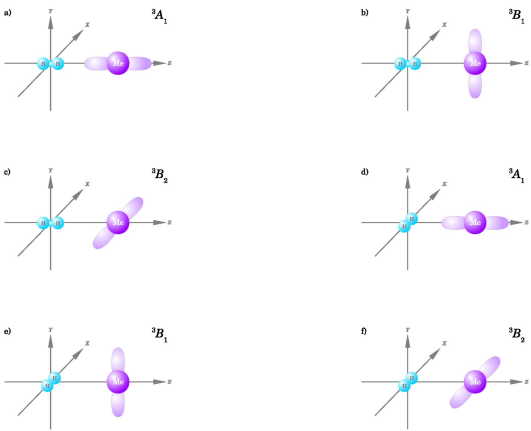}}}
\end{figure}

The calculations for the first excited Sr($^3P$)-H$_2$ and Yb($^3P$)-H$_2$ states involve considering three states corresponding to three possible spatial orientations of the singly occupied $p$-orbital and they can be classified according to their symmetry. Simplistic representations of the singly occupied $p$-orbital orientation in the first excited state are shown in Fig.~\ref{fig:orbit}. As the H$_2$ molecule approaches the Sr (Yb) atom collinearly ($C_{\infty v}$ symmetry), the degeneracy of the $^3P$ state is lifted and gives rise to one state of $^3A_1$ symmetry ($\Sigma$ state) and one doubly degenerate state of $^3B_1$ ($=$ $^3B_2$) symmetry ($\Pi$ state). In arrangements of lower symmetry, the latter state further splits into two states. In the collinear geometry ($C_{\infty v}$) the states can be labelled as $^3A_1$ and $^3B_1$ ($=$ $^3B_2$) -- see Figs.~\ref{fig:orbit}a-c. In the T-shape geometry ($C_{2v}$) we have the $^3A_1$, $^3B_1$ and $^3B_2$ states (\textit{cf.} Figs.~\ref{fig:orbit}d-f), whereas in bent symmetry ($C_s$) the states become $1^3A^{\prime}$, $2^3A^{\prime}$, and $^3A^{\prime\prime}$.

To ease the analysis, here we employ a nomenclature similar to that corresponding for the T-shape geometry, that is $A1$, $B1$ and $B2$. This approximately describes the orientation of the singly occupied $p$-orbital. In the case of the $A1$ surface the $p$-orbital is directed towards the hydrogen molecule (Figs.~\ref{fig:orbit}a and d), whilst in the case of the $B1$ surface the $p$-orbital is perpendicular to the complex symmetry plane (Figs.~\ref{fig:orbit}b and e). The $B2$ surface corresponds to the third situation, where the $p$-orbital is neither perpendicular to the complex symmetry plane nor directed towards H$_2$ (Figs.~\ref{fig:orbit}c and f).

\begin{figure}[ht]
\caption{The calculated intermolecular potential-energy curves for the ground (a) and the first excited (b--d) states of the Sr-H$_2$ complex}
\label{fig:surfs:srh2}
{\scalebox{0.35}{\includegraphics{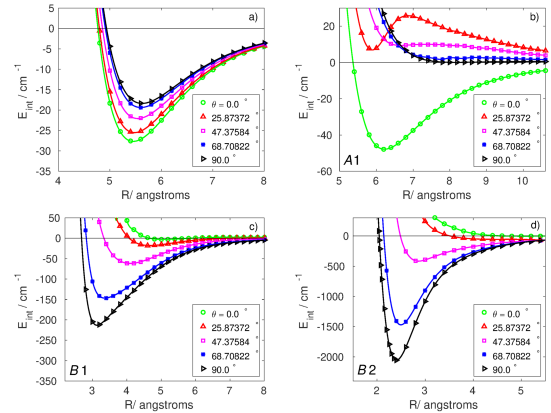}}}
\end{figure}

\begin{figure}[ht]
\caption{The calculated intermolecular potential-energy curves for the ground (a) and the first excited (b--d) states of the Yb-H$_2$ complex}
\label{fig:surfs:ybh2}
{\scalebox{0.35}{\includegraphics{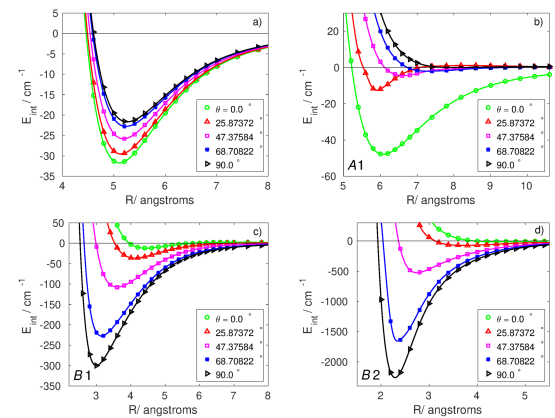}}}
\end{figure}

The calculated potential-energy curves for the ground and the first excited state of the Sr-H$_2$ and Yb-H$_2$ complexes are depicted  in Figs.~\ref{fig:surfs:srh2} and~\ref{fig:surfs:ybh2}, respectively. The values of the $\theta$ angle were chosen that they correspond to the abscissas of 9-points Lobatto-Gauss quadrature. In fact, because of the symmetry of the system, we needed to run the calculations only for 5 different angles.
The obtained interaction-energy values for all the studied complexes can be found in Supplementary Material.

Two minima and two transition states (equivalent by symmetry) are found on the ground-state IPESs. The global minimum is characterised by the collinear geometry, $R=$ 5.45~{\AA} (5.10~{\AA}), and an energy of $-$27.68~cm$^{-1}$ ($-$31.72~cm$^{-1}$), while the transition state corresponds to the T-shape geometry with $R=$ 5.62~{\AA} (5.26~{\AA}), and an energy of $-$18.37~cm$^{-1}$ ($-$21.65~cm$^{-1}$) in the case of the Sr-H$_2$ (Yb-H$_2$) complex. Analytical fits of both the ground-state IPESs along with Fortran subroutines for generating the potentials can be found in Supplementary Material.

The calculated excited-state surfaces (Figs.~\ref{fig:surfs:srh2}b-d and \ref{fig:surfs:ybh2}b-d) are deeper and more anisotropic than the respective ground-state one and they exhibit a similar pattern within both complexes.
In the following description the given value refers to the Sr-H$_2$ complex, while the value in parentheses to Yb-H$_2$.

The $A1$ surfaces (see Figs.~\ref{fig:surfs:srh2}b and \ref{fig:surfs:ybh2}b) are the shallowest ones with the global minimum of $-$47.83~cm$^{-1}$ ($-$47.84~cm$^{-1}$) for $\theta=0^{\circ}$ and $R=$ 6.22~{\AA} (6.08~{\AA}), but are still deeper than those of the ground state (\textit{cf.} Figs.~\ref{fig:surfs:srh2}a and \ref{fig:surfs:ybh2}a). As $\theta$ increases, the minima on the curves become shallower and for $\theta=90^{\circ}$ the potential energy curves have mainly repulsive character. Only a very shallow (less than 0.03~cm$^{-1}$ in absolute value) minimum (a transition state on the IPES) appears in the vicinity of $R=$ 8.02~{\AA} for Yb-H$_2$ (no minimum in the case of the Sr-H$_2$ system).

The $B1$ IPESs are much deeper than the $A1$ ones. The global minimum has an energy of $-$213.24~cm$^{-1}$ ($-$299.50~cm$^{-1}$) and corresponds to the T-shape geometry with $R=$ 3.13~{\AA} (2.99~{\AA}). The transition state at the collinear arrangement with an energy of $-$2.88~cm$^{-1}$ ($-$12.00~cm$^{-1}$) lies at $R=$ 5.05~{\AA} (4.44~{\AA}).

A similar pattern as the $B1$ IPESs present the $B2$ surfaces. However, these surfaces are the deepest one with the global minimum of ca. $-$2052.8~cm$^{-1}$ ($-$2260.1~cm$^{-1}$) in the T-shape geometry at $R=$ 2.41~{\AA} (2.29~{\AA}). The transition states are exactly the same as for the $B1$ surfaces, since both IPESs at the collinear arrangement correlate to the same state.

\subsubsection{Sr-He and Yb-He}
\label{sec:res:pess:he}
To check whether the above observed patterns in the calculated interaction energies reflects a more general trend, we compare our results with isoelectronic, but simpler complexes of Sr and Yb with He. Using the same methodology as above and analogous basis sets, we have calculated the potential-energy curves for both dimers. The aug-cc-pVTZ-DK and cc-pVTZ-DK bases for the He atom were used in the case of the Sr-He and Yb-He complexes, respectively.

The excited states are labelled similarly as for the complexes with H$_2$: $A1$ (the $^3A_1$ state) with the $p$-orbital along the Sr(Yb)-He interatomic axis ($\Sigma$ state) and $B1$ ($=$ $B2$) (the $^3B_1$ ($=$ $^3B_2$) states) with the $p$-orbital is perpendicular to the Sr(Yb)-He axis ($\Pi$ state) --- compare Fig.~\ref{fig:orbit}.

\begin{figure}[ht]
\caption{The calculated intermolecular potential-energy curves for the ground (a) and the first excited (b) states of the Sr-He and Yb-He complexes}
\label{fig:surfs:srhe+ybhe}
{\scalebox{0.35}{\includegraphics{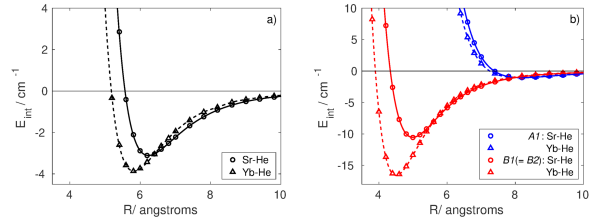}}}
\end{figure}

The results, depicted in Fig.~\ref{fig:surfs:srhe+ybhe}, are similar in both systems with the Yb-He curves being usually deeper than the Sr-He ones. The exception is the $A1$ state where the (weak) interaction (slightly exceeding 1~cm$^{-1}$ in absolute value) is comparable in both complexes.

In the ground state the interaction-energy minimum is $-$3.11~cm$^{-1}$ at 6.23~{\AA} for Sr-He and $-$3.87~cm$^{-1}$ at 5.82~{\AA} for Yb-He, while in the exited $B1$ state the minima are much deeper with energies of $-$10.54~cm$^{-1}$ (at 4.99~{\AA}) and $-$16.59~cm$^{-1}$ (at 4.50~{\AA}) in the Sr-He and Yb-He complexes, respectively.

The results reveal a similar trend as for the complexes with the hydrogen molecule and confirm a strong dependence of the interaction energy on the orientation of the singly occupied $p$-orbital: the interaction is attractive and much stronger when the $p$-orbital is perpendicular to the interatomic axis and is almost repulsive if the $p$-orbital is oriented towards the He atom.

\section{Summary}
\label{sec:sum}
Accurate IPESs for the ground and first excited states of Sr-H$_2$ and Yb-H$_2$ were constructed using a high-level \textit{ab initio} method and extensive basis sets. Methodological and basis-set studies were performed in order to ensure the proper choice of the approximation. 
In the calculations of the IPESs, the CCSD(T) method with the DKH Hamiltonian and the correlation-consistent basis sets of triple-$\zeta$ quality extended with 2 sets of diffuse functions and the 332211 midbond set were used.

The shapes of the ground-state IPESs are similar in both complexes, being relatively isotropic with two minima and two transition states (equivalent by symmetry). The global minima correspond to the collinear geometries with $R=$ 5.45 and 5.10~{\AA} and energies of $-$27.68 and $-$31.72~cm$^{-1}$ for the Sr-H$_2$ and Yb-H$_2$ systems, respectively.

The calculated excited-state surfaces for Sr($^3P$)-H$_2$ and Yb($^3P$)-H$_2$ are deeper and more anisotropic and, again, they exhibit a similar pattern within both complexes. The $A1$ surfaces are the shallowest ones with the global minimum of $-$47.83~cm$^{-1}$ ($-$47.84~cm$^{-1}$) in the collinear geometries for Sr-H$_2$ (Yb-H$_2$). The $B1$ IPESs are much deeper than the $A1$ ones, but the $B2$ surfaces are the deepest ones with the global minimuma of ca. $-$2052.8 and $-$2260.1~cm$^{-1}$ in the T-shape geometries at $R=$ 2.41 and 2.29~{\AA} for Sr-H$_2$ and Yb-H$_2$, respectively.

Additional calculations for the isoelectronic complexes of Sr and Yb with the He atom revealed a similar, strong dependence of the interaction on the orientation of the \textit{p}-orbital in the Sr($^3P$)-He and Yb($^3P$)-He states.

\vspace{0.4cm}

\textbf{\small Supplementary Material}

Analytical fits of the ground-state IPESs. Complete tables for the basis set studies. The calculated interaction energies for the Sr-H$_2$, Yb-H$_2$, Sr-He, and Yb-He complexes. Fortran subroutines for generating the Sr-H$_2$ and Yb-H$_2$ ground-state potentials.

\begin{acknowledgments}
Research financed by the National Science Centre in Poland within the OPUS 8 project No. 2014/15/B/ST4/04551.
Support has been received from the project \textit{EMPIR 15SIB03~OC18}. This project has received funding from the EMPIR programme co-financed by the Participating States and from the European Union’s Horizon 2020 research and innovation programme.
Calculations have been carried out using resources provided by Wroclaw Centre for Networking and Supercomputing (\url{http://wcss.pl}), Grant No. 294. The research is a part of the program of the National Laboratory FAMO in Toru{\'n}, Poland.
\end{acknowledgments}

\end{document}